\documentstyle[aps,prl,preprint]{revtex}
\begin{document}
\draft
\title{Nucleosynthesis and the Mass of the $\tau$ Neutrino}
\author{Steen Hannestad and Jes Madsen}
\address{Institute of Physics and Astronomy,
University of Aarhus, 
DK-8000 \AA rhus C, Denmark}
\date{November 22, 1995}
\maketitle

\begin{abstract}
The primordial abundance of long-lived heavy Majorana neutrinos 
is calculated from the full Boltzmann equation.
Inclusion of scattering reactions
drastically change the predicted abundance of a heavy
neutrino species. This loosens the well known mass
constraint on MeV neutrinos from Big Bang nucleosynthesis,
and allows for the existence of a Majorana $\tau$ neutrino with
mass $m_{\nu_{\tau}} \geq 11$ MeV.
Further experimental efforts are therefore
needed to investigate the range
$11 \text{MeV} \leq m_{\nu_{\tau}} \leq 24 \text{MeV}$.
Some interesting cosmological consequences of an MeV $\nu_\tau$
are also pointed out.
\end{abstract}

\pacs{98.80.Ft, 14.60.Pq, 26.35.+c\\ \\ \\ Physical Review Letters (in press)}

The possibility of heavy $\tau$ neutrinos in the MeV region has
been investigated many times in the literature,
primarily because such a neutrino could have
very interesting consequences for both cosmology and
supernovae \cite{turner}.
The best current experimental limit to the $\tau$ neutrino mass comes
from the ALEPH collaboration \cite{buskulic}, 
and is $m_{\nu_{\tau}} \leq 24$ MeV with 95 \% CL.
It is well known that primordial nucleosynthesis puts
stringent limits on the mass and lifetime of the
$\tau$ neutrino 
\cite{dicus,scherrer,turnerII,dolgov,dodelson,kolb,kawasaki}. 
These bounds are most stringent for neutrinos in the MeV
region. This is because such a neutrino decouples at
a temperature of a few MeV, where its abundance is
still comparable to the density of the massless species.
But when the temperature drops, the energy density of
the massive neutrinos grows relative to that
of massless particles. For low temperatures, the
energy density of an MeV neutrino will be many times
that of the other particles present, effectively making
the universe mass dominated, and significantly changing the
outcome of nucleosynthesis. 
If the mass is much higher,
the neutrinos will have been Boltzmann suppressed at
the time they decouple, to a level where they are not significant
during nucleosynthesis.

The most recent investigations of the effects of
a massive Majorana $\tau$-neutrino in the MeV region on nucleosynthesis,
are those of Kolb et al. \cite{kolb} and Kawasaki et al. 
\cite{kawasaki}. For a standard Majorana neutrino that is stable
on the timescale of nucleosynthesis Kolb et al. find an excluded
mass interval of 0.5-32 MeV, whereas Kawasaki et al. find
an exclusion interval of 0.1-50 MeV. Combined with the experimental data
this means that a massive $\tau$ neutrino with mass greater
than 0.1-0.5 MeV and lifetime greater than $10^{3}$ s is
excluded by nucleosynthesis.
However, most of these calculations, except \cite{kawasaki}, use the integrated
Boltzmann equation. Furthermore, they assume that the
distribution function is kept in kinetic equilibrium
at all times by scattering reactions \cite{bernstein,gondolo}.
Kawasaki et al. \cite{kawasaki} have used the full equation, but 
assuming only annihilation interactions, no scattering.
Furthermore they use Boltzmann statistics.
It is therefore important to see how these limits change
when we use the full Boltzmann equation, with all possible
interactions.
Below we calculate the relic abundance of a massive
Majorana $\tau$ neutrino with lifetime longer than the time of 
nucleosynthesis ($\tau \geq 10^{3}$ s). 
We then use nucleosynthesis to put limits on the allowed mass range, and
find that a significant window for $m_{\nu_{\tau}}\geq 11$ MeV is still open.

The fundamental equation used is the Boltzmann equation
that describes the evolution of a particle species. 
We assume that all particle distributions are homogeneous in space
and isotropic in phase-space. In that case the 
Boltzmann equation can be written as 
\begin{equation}
C_{\text{L}}[f] = \sum C_{\text{coll}}[f],
\end{equation}
where $C_{\text{L}}[f] = {\partial f}/{\partial t}-
Hp{\partial f}/{\partial p}$
is the Liouville operator, and
\(\sum C_{\text{coll}}\) is the sum of all possible collisional 
interactions.
We follow Ref. \cite{hannestad} in assuming that only 
two-particle interactions are important.
The collision operator can then be written as
\begin{eqnarray}
C_{\text{coll}}[f] & = & \frac{1}{2E_{1}}\int d^{3}\tilde{p}_{2}
d^{3}\tilde{p}_{3}d^{3}\tilde{p}_{4}
\Lambda(f_{1},f_{2},f_{3},f_{4})\times 
\label{integral}\\ 
& & S \mid \! M \! \mid^{2}_{12\rightarrow 34}\delta^{4}
({\it p}_{1}+{\it p}_{2}+{\it p}_{3}-{\it p}_{4})(2\pi)^{4}, 
\nonumber
\end{eqnarray}
where \(\Lambda(f_{1},f_{2},f_{3},f_{4})=(1-f_{1})(1-f_{2})f_{3}
f_{4}-(1-f_{3})(1-f_{4})f_{1}f_{2}\) is the phase space factor, 
including Pauli blocking of the final states, and $d^{3}\tilde{p}
 = d^{3}p/((2 \pi)^{3} 2 E)$. $S$ is a symmetrization
factor of 1/2! for each pair of identical particles in initial or 
final states,
and $\mid\!\!M\!\!\mid^{2}$ is the weak interaction matrix element
squared, and appropriately spin summed and averaged.
${\it p}_{i}$ is the four-momentum of particle $i$.
Details of how to solve this equation, as well as the
relevant matrixelements, are given in Ref.
\cite{hannestad}. 
In addition to the Boltzmann equation we
need equations relating time, temperature and the
expansion rate, $H$.
These will be supplied by the energy conservation 
equation
${d}(\rho R^{3})/dt + p {d}(R^{3})/dt = 0$,
and the Friedmann equation $H^{2} = 8 \pi G \rho/3$.
It is assumed, at all times, that $e^{\pm}$ are kept in thermodynamic
equilibrium with the photon gas by electromagnetic
interactions.
We parametrize the momenta of all particles in terms of the
parameter $z = p_{\nu}/T_{\gamma}$, and use
a grid of 25 points to cover the range in $z$.
The system of equations is then evolved in time by a Runge-Kutta
integrator, giving us the distribution function of each
particle species, as well as the scale factor $R$, as a function 
of photon temperature.
By doing it this way, we have the advantage that we know
the specific distribution functions for all neutrino
species, whereas
a treatment using the integrated Boltzmann equation has to make
an assumption of kinetic equilibrium, giving only a rough approximation
to the true distribution functions (in fact we find that the
distributions can deviate substantially from kinetic equilibrium).

Fig.\ \ref{fig1}
shows the final relic density of $\nu_\tau$ before
decay becomes important. $rm_{\nu}$ is defined as 
$rm_{\nu}=m_{\nu}n_{\nu}/n_{\nu}(m=0)$. 
We have compared our results to those of Kolb et al. \cite{kolb},
who use the integrated Boltzmann equation.
For small masses, it is not
important whether we use the integrated or the full 
Boltzmann equation. The reason is that the neutrino decouples
before annihilation really commences, therefore the treatment
of the annihilation terms is not very important. 
For masses larger than 5 MeV there is, however, a significant
difference, increasing with mass. 
This is an effect of large deviations from kinetic equilibrium.
The cross-sections increase
with energy. Thus the high momentum states are
depleted relatively more than low momentum states, only partly
compensated by upscattering. For $m_\nu \gg T$ the 
thermally averaged annihilation cross section, 
$< \! \sigma \! \mid \! v \! \mid>$, is larger
than the corresponding value calculated for a kinetic equilibrium
distribution, as normally used in the integrated Boltzmann equation.
This is the main reason why the final abundance of $\nu_\tau$ is lower if
we use the true Boltzmann equation rather than the integrated one.

In Fig.\ \ref{fig2} we show the final distribution of a 10 MeV $\nu_\tau$.
The difference between the distributions with and without
scattering interactions is striking. 
If only annihilations are taken into account, the high momentum states
are almost empty, because they cannot be refilled. On the other hand,
the low momentum states are highly populated, because they
annihilate more slowly and cannot be scattered away.
The result of this effect is clearly seen in Fig.\ \ref{fig1}.
For large
values of $m_{\nu_{\tau}}$, the number density is much higher
if we use only annihilation interactions, because once the
high momentum states are depleted, almost no annihilation will
take place. Fig.\ \ref{fig2} also shows the very significant deviation
between the actual distribution and a kinetic
equilibrium distribution, $f_{\nu} = 1/(e^{(E-\mu)/T}+1)$, 
with the same number density.

Fig.\ \ref{fig3} shows the final distribution of the electron
neutrino. It is seen that the distribution functions can
become highly non-thermal, because the pairs created by
annihilation of $\tau$ neutrinos cannot thermalize sufficiently
before the electron neutrinos decouple completely.
This non-thermality can have a significant effect on the
production of light elements, especially $^{4} \text{He}$, because
the n-p converting reactions depend not only on the energy
density of neutrinos, but also on the specific form
of the distribution function for $\nu_e$, which enters in the n-p
conversion rate integrals. Overall, the increase in the cosmic expansion
rate due to the extra energy density in $\tau$- (and massless)
neutrinos, which increases $^4\text{He}$-production, is to some extent
compensated by a decrease in neutron-fraction due to the change in the
n-p rate integrals. 

We also note that the final
number densities of muon and electron neutrinos exceed the result for 
the standard scenario with 3 low-mass flavors by a significant amount.
For eV-mass $\nu_\mu$ or $\nu_e$ this changes the present day
contribution to the cosmic density to $\Omega_\nu h^2=\alpha m_\nu
/93.03$eV  with $\alpha =1$ for a massless $\nu_\tau$, and $\alpha
=$1.07(1.10), 1.07(1.15), 1.09(1.24), 1.16(1.51) for $\nu_e$($\nu_\mu$)
for $m_{\nu_\tau}=$5, 10, 15, 20 MeV ($h$ is the Hubble-parameter in
units of 100km s$^{-1}$ Mpc$^{-1}$). Later decay of $\nu_\tau$ can
further increase the value of $\alpha$. This may have significant
implications for galaxy formation scenarios.

In order to derive mass limits on $\nu_\tau$ we need to
calculate the predicted abundances of the different light 
elements and compare them with observations. 
To do this, we have changed the nucleosynthesis code of
Kawano \cite{kawano} to incorporate a massive 
Majorana $\tau$ neutrino. 
As previously mentioned, we assume that decay has no effect during
nucleosynthesis ($\tau \geq 10^{3} \text{s}$).
Note that we not only have to change the energy density of the
$\tau$ neutrino, but also the energy density in massless neutrinos,
as well as the specific distribution function, $f_{\nu_{e}}$, of
electron neutrinos in the nucleosynthesis code.

Fig.\ \ref{fig4} shows the relic abundances of the different light 
elements as
a function of the baryon-to-photon ratio, $\eta$, for different
$m_{\nu_\tau}$.

The observational limits on the primordial $^{4} \text{He}$ 
abundance have been evaluated by Olive and Steigman \cite{olive} to be
$Y_{P} = 0.232 \pm 0.003 \pm 0.005$,
where the first uncertainty is a 1 $\sigma$ statistical uncertainty,
and the second an estimated 1 $\sigma$ systematic uncertainty. 
However, the actual systematic
uncertainty might be substantially larger than this.
Copi et al. \cite{copi} quote a systematic uncertainty of 
$+0.016/-0.012$ as more realistic. 
Furthermore there are some indications that the $^{4} \text{He}$ abundance
has been systematically underestimated, and might really be as large
as 0.255 \cite{sasselov}.

A lower limit to the primordial deuterium abundance has been 
calculated by Hata et al. \cite{hata}, from interstellar medium data,
 to be $\text{D/H} \geq 1.6 \times 10^{-5}$. The recent observation of an 
apparently very high deuterium abundance in QSO absorption 
systems would imply an upper limit of around $2 \times 10^{-4}$
\cite{carswell,songaila,schramm}.
The interpretation of these observations is, however, uncertain
\cite{steigman}.
Normally, one uses an upper limit on $\text{D} + ^{3}\text{He}$ 
instead of D alone.
This gives $(\text{D} + ^{3}\text{He)/H} \leq 1.1 \times 10^{-4}$
\cite{copi} (again from local galactic data), incompatible
with the high deuterium value. The upper limit from this method
is unfortunately also very uncertain, because it depends on
evolution effects of $^{3}\text{He}$, 
which are not particularly well known \cite{schramm}.
Finally the abundance of $^{7}\text{Li}$ is also used. Copi et al.\cite{copi}
use a limit of $^{7}\text{Li/H} = 1.4 \pm 0.3 ^{+1.8}_{-0.4} \times 10^{-10}$.  

Based on these data we use a strong limit to the primordial
abundances of $Y_{P} = 0.232 \pm 0.006$, $\text{D/H} \geq 1.6 \times 10^{-5}$,
$(\text{D} + ^{3}\text{He)/H} \leq 1.1 \times 10^{-4}$ and 
$^{7}\text{Li/H} = 1.4^{+1.8}_{-0.5} \times 10^{-10}$. Furthermore we
use a weak (and perhaps rather more realistic) limit of
$Y_{P} = 0.232^{+0.016}_{-0.012}$,
$\text{D/H} \geq 1.6 \times 10^{-5}$,   
$\text{D/H} \leq 2 \times 10^{-4}$ and 
$^{7}\text{Li/H} = 1.4^{+1.8}_{-0.5} \times 10^{-10}$.

We now use these limits to infer mass limits on the $\tau$ neutrino.
Using the strong limit we obtain a minimum allowed mass of 20 MeV, 
and using the weak limit we obtain a minimum mass of 11 MeV. 
Thus, in contrast to the conclusion of earlier work, there is still
an allowed mass interval of 11--20 MeV $\leq m_{\nu_{\tau}}
\leq$ 24 MeV\cite{deltannu}.
These limits all apply to Majorana neutrinos. Similar effects will occur
for Dirac neutrinos, also opening a new mass window for them, but the
calculations with the full Boltzmann equations are complicated by the
need to include spin-flip interactions.

Our lower limits on the $\tau$ neutrino mass are much weaker than
those of Kolb et al. and Kawasaki et al. As discussed previously,
this is to be expected since Kolb et al. use the integrated
Boltzmann equation and Kawasaki et al. only consider annihilation
interactions.
Thus the conclusions of earlier works, that an MeV
$\tau$ neutrino is excluded by nucleosynthesis,
is changed if we use the full Boltzmann equation with all possible
interactions.
New experiments to directly search for a neutrino in this mass
range are therefore needed. 
Furthermore, if the true $^{4}$He abundance turns out to be high,
as pointed to by some authors \cite{sasselov}, 
and the baryon density low,
as suggested by the lack of microlensing events in the galactic 
halo, a heavy $\tau$ neutrino could be a natural way to bring
the predictions of Big Bang nucleosynthesis into consistency 
with observations.

\acknowledgments
This work was supported in part by the Theoretical Astrophysics Center
under the Danish National Research Foundation, and by the European Human
Capital and Mobility Program.

\begin{figure}
\caption{The relic number density times mass,
$r m_{\nu}$, for massive $\tau$ 
neutrinos. The solid curve is calculated using the full
Boltzmann equation and all interactions. The dashed curve
is obtained by using the full equation, but with only annihilation
interactions. The dotted curve is adopted from 
Kolb et al.\protect\cite{kolb}}
\label{fig1}
\end{figure}

\begin{figure}
\caption{The distribution of $\tau$ neutrinos
of mass 10 MeV at a photon temperature of 0.03 MeV.
The solid line comes from using all possible interactions and the
dotted curve is from using only annihilations. The dashed curve is a
kinetic equilibrium distribution with the same number density 
as that of the solid curve. Notice that the actual
distribution is far from kinetic equilibrium.}
\label{fig2}
\end{figure}

\begin{figure}
\caption{The electron neutrino distribution
at a photon temperature of 0.03 MeV.
The solid curve is for $m_{\nu_{\tau}} = 5$ MeV, the dotted for
$m_{\nu_{\tau}} = 10$ MeV and the dashed for $m_{\nu_{\tau}} = 15$ MeV.
For comparison, a standard $\nu_e$-distribution without input from
$\nu_\tau$-annihilations is shown as the dot-dashed curve.}
\label{fig3}
\end{figure}

\begin{figure}
\caption{Predicted light element abundances as a function of
the baryon-to-photon ratio $\eta_{10} = 10^{10} n_{B}/n_{\gamma}$
for different $\tau$ neutrino masses.
The solid curves are for $m_{\nu_{\tau}} = 10$ MeV, the dotted for
$m_{\nu_{\tau}} = 15$ MeV and the dashed for $m_{\nu_{\tau}} = 20$ MeV.
The dot-dashed curves are the predictions for $m_{\nu_{\tau}} \rightarrow 
\infty$, corresponding to 2 massless neutrinos.
Horizontal lines are observational limits, the dot-dot-dot-dashed
lines correspond to the strong limit and the long-dashed to the
weak limit. Note that in some cases the strong and weak limits
coincide.}
\label{fig4}
\end{figure}


\begin{references}
\bibitem{turner}For recent reviews, see G. Guyk and M. S. Turner, 
Nucl. \ Phys. \ B \ Proc. \ Suppl. {\bf 38}, 13 (1995);
G. Gelmini and E. Roulet, Rep. Prog. Phys
{\bf 58}, 1207 (1995).
\bibitem{buskulic}D. Buskulic et al., Phys. Lett. B {\bf 349},
585 (1995).
\bibitem{dicus}D. A. Dicus et al., Phys.\ Rev.\ D 
{\bf 17}, 1529 (1978).
\bibitem{scherrer}E. W. Kolb and R. J. Scherrer, Phys.\ Rev.\ D
{\bf 25}, 1481 (1982).
\bibitem{turnerII}R. J. Scherrer and M. J. Turner, Ap. J. 
{\bf 331}, 19 (1988).
\bibitem{dolgov}A. D. Dolgov, K. Kainulainen and I. Z. Rothstein,
Phys.\ Rev.\ D {\bf 51}, 4129 (1995).
\bibitem{dodelson}S. Dodelson, G. Guyk and M. S. Turner,
 Phys.\ Rev.\ D {\bf 49}, 5068 (1994).
\bibitem{kolb}E. W. Kolb et al., Phys.\ Rev.\ Lett.
{\bf 67}, 533 (1991); see also A. D. Dolgov and I. Z. Rothstein, {\it
ibid\/} {\bf 71}, 476 (1993).
\bibitem{kawasaki}M. Kawasaki et al., Nucl. Phys. B,
{\bf 419}, 105 (1994).
\bibitem{bernstein}J. Bernstein, L. S. Brown and G. Feinberg,
Phys.\ Rev.\ D {\bf 32}, 3261 (1985).
\bibitem{gondolo}P. Gondolo and G. Gelmini, Nucl. \ Phys. \ B \
{\bf 360}, 145 (1991).
\bibitem{hannestad}S. Hannestad and J. Madsen, Phys.\ Rev.\ D
{\bf 52}, 1764 (1995).
\bibitem{kawano} L. Kawano, Report No. FERMILAB-Pub-92/04-A (1992)
(unpublished).
\bibitem{olive}K. A. Olive and G. Steigman, Ap. J. Suppl.
{\bf 97}, 49 (1995).
\bibitem{copi}C. J. Copi, D. N. Schramm and M. S. Turner,
Phys.\ Rev.\ Lett.\ {\bf 75}, 3981 (1995).
\bibitem{sasselov}D. Sasselov and D. Goldwirth, Ap. J. Lett.
{\bf 444}, L5 (1995).
\bibitem{hata}N. Hata et al., Phys.\ Rev.\ Lett.\ {\bf 75}, 3977 (1995).
\bibitem{carswell}R. F. Carswell et al., MNRAS {\bf 268}, L1 (1994).
\bibitem{songaila}A. Songaila et al., Nature {\bf 368}, 599 (1994).
\bibitem{schramm}D. N. Schramm, in {\it Proceedings of the ESO/EIPC
Workshop on The Light Element Abundances}, edited by P. Crane (Springer,
Berlin, 1995), p. 51.
\bibitem{steigman}G. Steigman, MNRAS {\bf 269}, L53 (1994).
\bibitem{deltannu}Another standard manner of expressing nucleosynthesis
results is in terms of the equivalent number of massless neutrinos,
$N_\nu$, because the $^4$He-production is roughly given by $Y_P\approx
0.2253+0.011\ln\eta_{10}+0.0139(N_\nu-3)$. We find that $N_\nu$ takes
the values 3.80, 3.47, 2.90, and 2.00 for $m_{\nu_\tau}$ of
10, 15, 20 MeV and $\infty$. Using $N_\nu\leq 3.4$ Kolb et
al.\cite{kolb} obtain a bound of 25 MeV (above the current experimental
limit), which for our results is reduced to 16 MeV (well below the
experimental limit).
\end{references}
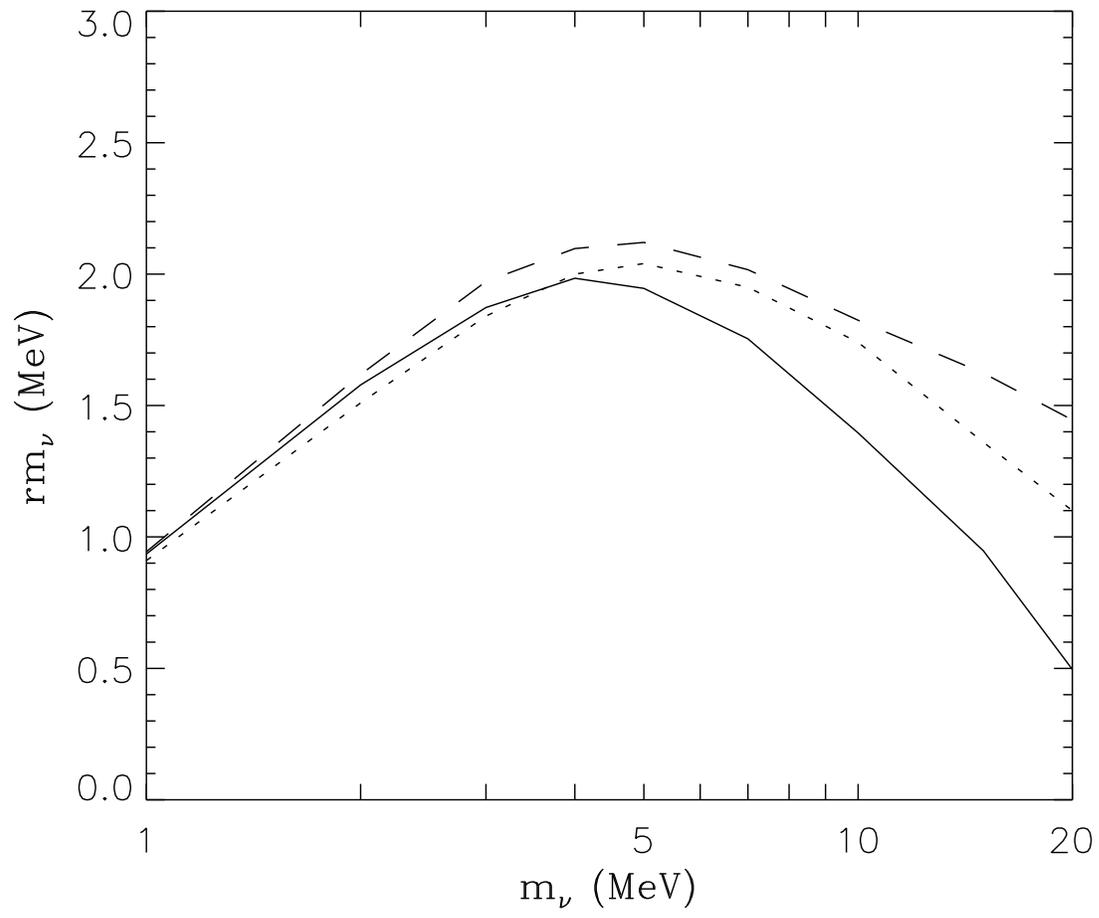
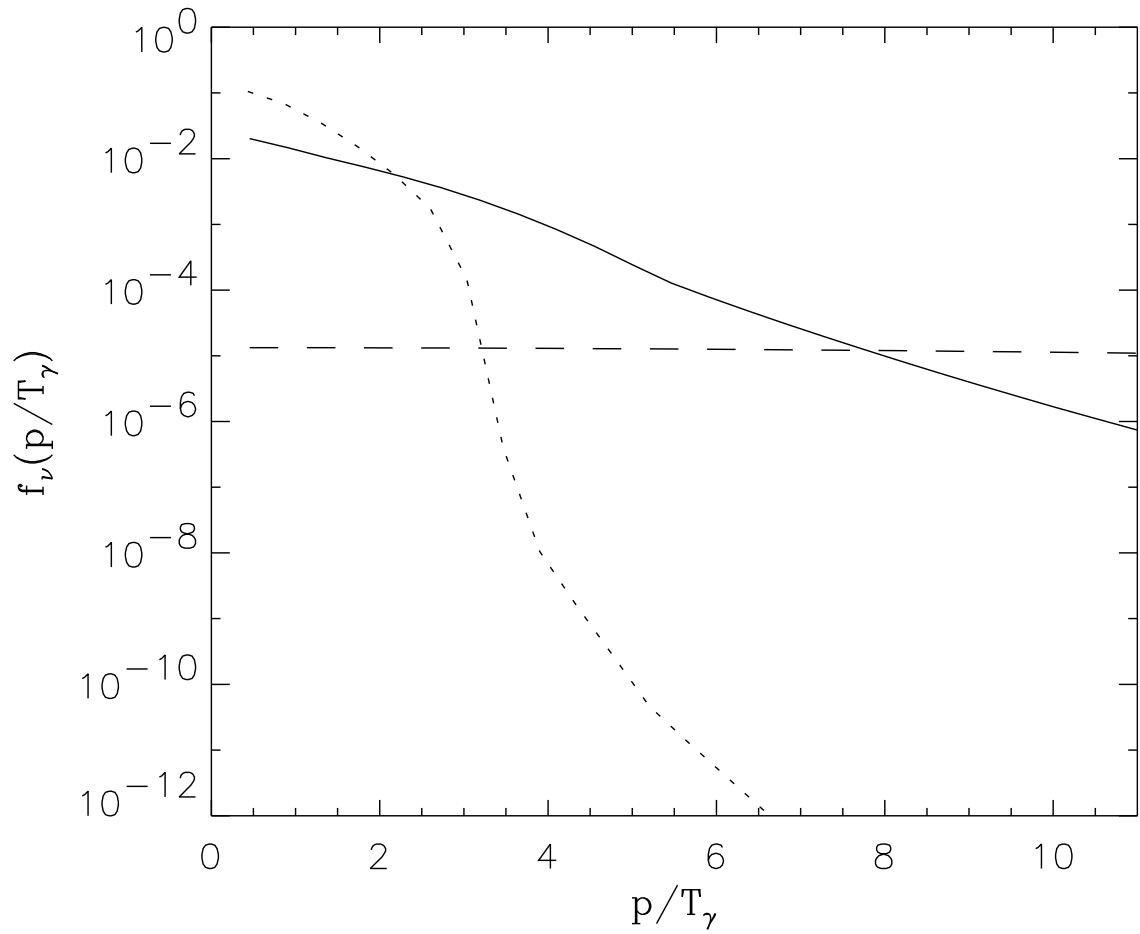
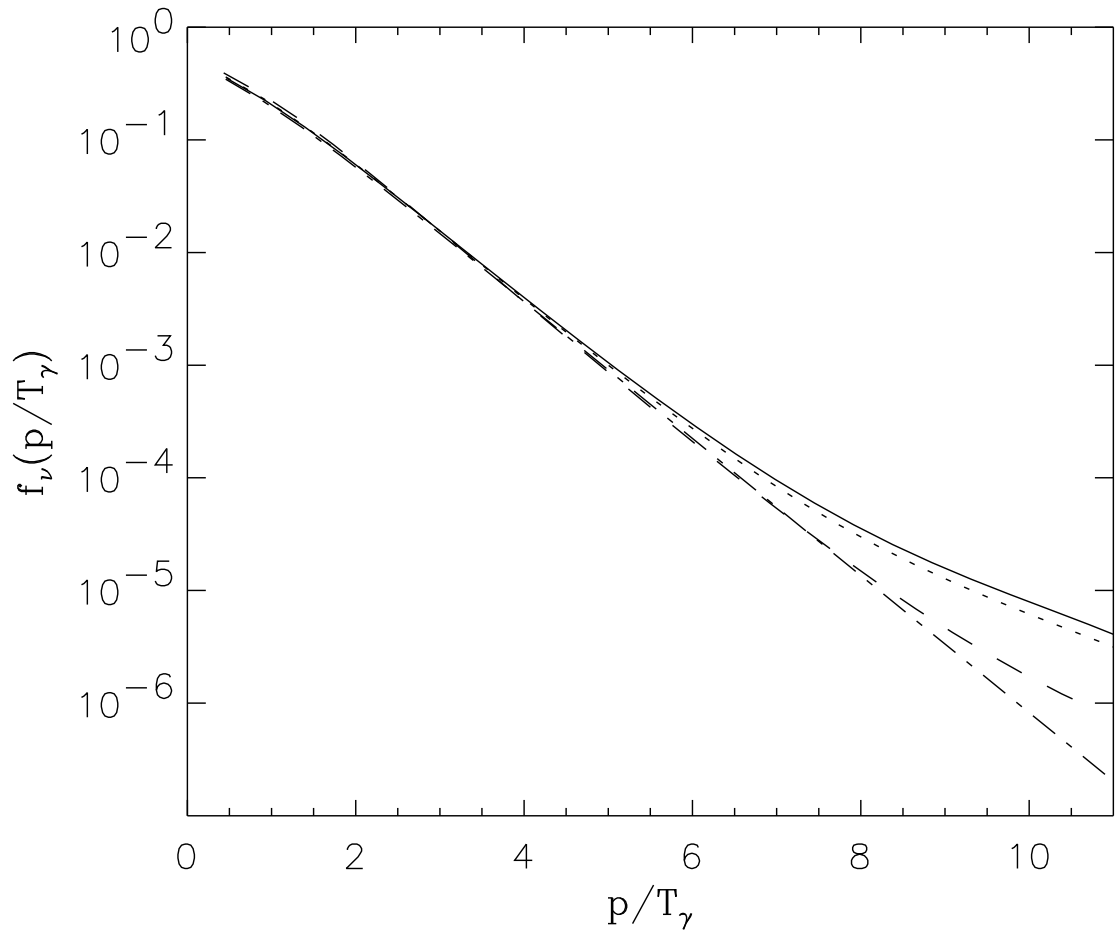
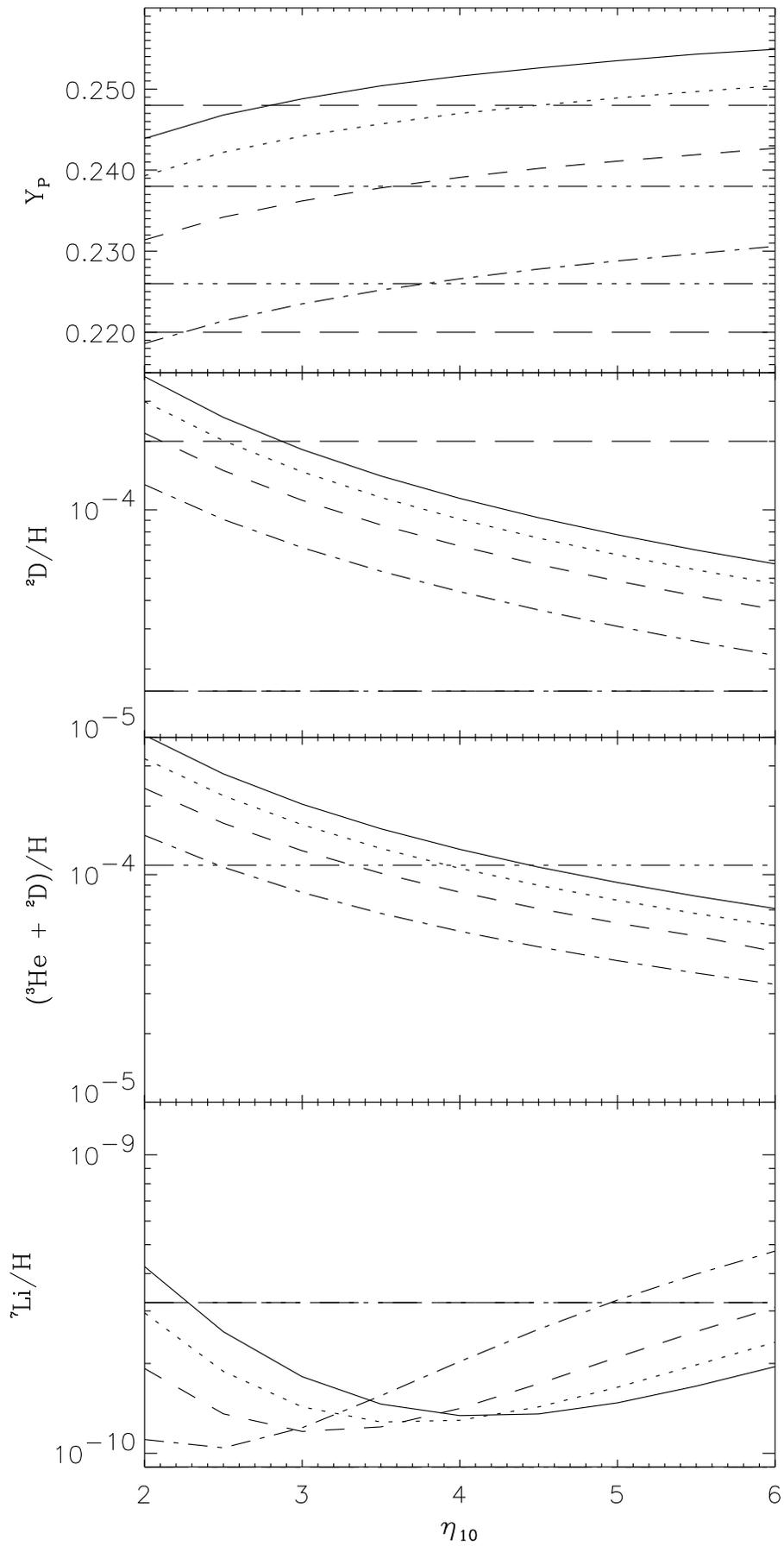
\end{document}